# Graphene based Supercapacitors with Improved Specific Capacitance and Fast Charging Time at High Current Density

Santhakumar Kannappan[a], Karthikeyan Kaliyappan[b,c], Rajesh Kumar Manian[d], Amaresh Samuthira Pandian[b], Hao Yang[e], Yun Sung Lee[b], Jae-Hyung Jang[a,f] and Wu Lu[a,e]*

a) Department of Nanobio Materials and Electronics, Gwangju Institute of Science and Technology, Gwangju 500-712, Republic of Korea.
b) Faculty of Applied Chemical Engineering, Chonnam National University, Gwangju, 500-757, Republic of Korea.
c) Department of Mechanical and Materials Engineering, The University of Western Ontario, London, Ontario, N6A 5B9, Canada
d) Department of Chemistry, Institute of Basic Science, Chonnam National University, Gwangju, 500-757, Republic of Korea
e) Department of Electrical and Computer Engineering, The Ohio State University, Columbus, OH 43210, USA.
f) School of Information and Communications, Gwangju Institute of Science and Technology, Gwangju 500-712, Republic of Korea.

*To whom all correspondence should be addressed. E-mail address: lu@ece.osu.edu

Graphene is a promising material for energy storage, especially for high performance supercapacitors. For real time high power applications, it is critical to have high specific capacitance with fast charging time at high current density. Using a modified Hummer's method and tip sonication for graphene synthesis, here we show graphene-based supercapacitors with high stability and significantly-improved electrical double layer capacitance and energy density with fast charging and discharging time at a high current density, due to enhanced ionic electrolyte accessibility in deeper regions. The discharge capacitance and energy density values, 195 $Fg^{-1}$ and 83.4 $Whkg^{-1}$, are achieved at a current density of 2.5 $Ag^{-1}$. The time required to discharge 64.18 $Whkg^{-1}$ at 5 A/g is around 25 sec. At 7.5 $Ag^{-1}$ current density, the cell can deliver a specific capacitance of about 137 $Fg^{-1}$ and maintain 98 % of its initial value after 10,000 cycles, suggesting that the stable performance of supercapacitors at high current rates is suitable for fast charging-discharging applications. We attribute this superior performance to the highly porous nature of graphene prepared with minimum restacking due to crimple nature wrinkles and the improved current collecting method.

**Introduction**

Graphene and related materials have been considered as potential functional materials in industrial applications such as electrical**,** electrochemistry and electronic applications.[1-3] Large volume energy storage supercapacitors with high power density, low manufacturing cost and low maintenance cost is a key research aspect to be addressed in energy storage systems. Recently, graphene has been a promising material in battery research for energy storage with high power density and long cycle life.[4] Beside its light weight with high surface area, carbon is more readily available material in abundant in nature. Though supercapacitors are not as good as the existing battery devices in term of energy density, further research in this area could lead to graphene based supercapacitors with improved energy density that is comparable with lithium ion batteries but with a higher power density and a rapid charging system. The charge storage in supercapacitors is based on electrochemical double layer capacitance i.e. formation of interfacial double layer on active materials.[5] There are several reports on graphene based supercapacitors, in recent, to improve specific capacitance using various electrolytes like potassium hydroxide, organic electrolyte and ionic liquids.[6-11] Chenguang et al[4] reported graphene based supercapacitor with high energy density. This was achieved for a high surface area graphene with curved morphology. However, their reported value (154.1 $Fg^{-1}$) of discharge capacitance at 1 A/g is still relatively low and the discharge time is 2 minutes at an energy density of 76.3$Whkg^{-1}$. Later, Yanwu et al[9] have investigated the capacitance properties of the graphene supercapacitors with the same ionic liquid. However, the specific capacitance reported for their devices is still low and much improvement has to be made in graphene synthesis and current collecting method. Further, many research groups have reported pseudocapacitors with metal oxide and polymer graphene composites with improved performance which involve Faradic reactions.[10-16] But the integrity and mechanical stability of the electrode is poor in such pseudocapacitors during electrochemical reactions, leading to an expansion in the polymer based electrode region that has to be addressed before it comes to real time applications.[10-12] Metal oxide graphene composites that have been used to improve the capacitance performance include MgO, $Fe_3O_4$, $MnO_2$, and cobalt oxide.[12-16] Though these metal oxide composites have shown improvements on the specific capacitance and energy density but the charging time and discharging time are still too low for a high power application where the energy delivery should be fast to various load applications. Further improvement in capacitance has been achieved by using various ionic liquids. 1-Ethyl-3-methylimidazolium tetrafluoroborate (EMIBF) has shown an improved capacitance and moreover, the operating voltage is raised to 3.5 V compared to aqueous electrolyte (1.2 V).[9]

The key of graphene synthesis for high performance supercapacitors is to have a high surface area and good electrical activation.[17] There are several methods to synthesis graphene such as Hummer's method[18-21], dispersion method[22], microwave method[23,12], and electrochemical method[24]. But, the graphene synthesized from the above mentioned methods still delivered low capacitance performance, which cannot be adopted for practical applications. It is well known that the electrochemical performance of graphene could be enhanced by improving their surface and morphological properties. Graphene synthesis method has to be improved on the oxidiation, exfoliation and reduction process. In this work we synthesized graphene using tip sonication for exfoliation based on the modified Hummer's method.[25] Using ionic liquid a superior capacitance value with an operating voltage of 3.5 V was achieved with energy density 83.36 Whkg$^{-1}$ at a current density of 2.5 Ag$^{-1}$ at room temperature. Several cells were assembled to check the repeatability of the performance and all cells exhibited similar results with plus or minus 15 % variation in specific capacitance.

**Experimental**

**Synthesis of graphene few layers**

Graphene oxide (GO) was prepared from graphite powder by the modified Hummers method.[25] In brief, the graphite powder (4 g) was first preoxidized with a solution of concentrated $H_2SO_4$ (60 mL), $K_2S_2O_8$ (2 g), and $P_2O_5$ (2 g) at 80 °C. The resulting mixture, after cooling to room temperature, was filtered and washed until the rinse water pH became neutral.

The oxidized graphite powder (2 g) was placed in cold (0 °C) concentrated $H_2SO_4$ (40 mL), and $KMnO_4$ (6 g) was added subsequently under stirring in an ice-bath. The mixture was then stirred at 35 °C for 2 h, after which DI water (92 mL) was added. Next, additional DI water (280 mL) and 30% $H_2O_2$ solution (20 mL) were added to the mixture to stop the reaction. The resulting mixture was washed by repeated centrifugation and filtration with 5% HCl solution in order to remove the metal ions and washed with DI water till the pH value became neutral. Finally, drying under vacuum at 50 °C for 6 hours afforded the GO product. GO (500 mg) and DI water (100 mL) were added into a beaker and was sonicated using tip sonicator (750 Watt Ultrasonic Processor, Sonics) for 20 min with 20% power. A microprocessor based and programmable ultrasonic processor was used (Sonics and Materials Inc., Model: VCX 750) at a frequency of 20 kHz. GO was sonicated to exfoliate the suspension completely and centrifuged at 4000 rpm to remove the unexfoliated GO. The GO reduction was carried out similar to a procedure reported by Li et al.[4] In a typical synthesis procedure, about 8.1 mL of the purified exfoliated GO solution was diluted to 90 mL using distilled water. Into this solution, 50 μL of hydrazine solution (35 wt% in water) was added and stirred for 1 min.

**Characterization**

The graphene synthesized were characterized by various surface, structural and compositional analyzing techniques. Graphene hexagonal peak were identified by high resolution X-ray diffractometer (Rigaku, Japan) with Cu Ka radiation (k = 1.54056 Å). Raman spectra were obtained with a Horiba Jobin-Yvon, France using 514 nm $Ar^+$ ion laser as the excitation source with 10 mW power on the sample surface and a resolution around 0.5 $cm^{-1}$. X-ray photoelectron spectroscopy, MULTILAB 2000 SYSTEM, SSK, U.S.A was performed for chemical analysis. The surface morphological measurements were carried out by field emission scanning electron microscopy (FESEM) using S-4700 HITACHI, Japan and high-resolution transmission electron microscopy (HRTEM) using TECNAI F20 [Philips], respectively. Fourier transform infrared (FTIR) study was also carried out to examine the vibrational characteristics of graphene using an IRPrestige-21, Shimadzu, Japan. The surface area and porosity of the synthesized powders was determined by Brunauer−Emmett−Teller (BET) adsorption method and Barrett-Joyner-Halenda (BHJ) method respectively using low temperature nitrogen adsorption surface area analyzer (ASAP 2020, Micromeritics Ins, USA).

**Fabrication of Coin cells and electrochemical testing**

The working electrode was fabricated with 75 wt% of graphene powder, 18 wt% of Ketjen black and 7 wt% of teflonized acetylene black (TAB), which was pressed on a stainless steel (SS) mesh, under a pressure of 300 kg/cm$^2$ and dried at 140 °C for 5 h in an oven. The electrochemical measurements were performed using coin-type CR2032 cells. A porous polypropylene film (Celgard 3401) was used as the separator. The test cell was fabricated in an argon filled glove box by pressing together the graphene electrodes separated by the separator. 1-butyl-3-methylimidazolium tetrafluoroborate (BMIM BF$_4$) ionic liquid was used as the electrolyte.

Galvanostatic charge-discharge measurements of the cell were carried out between 0-3.5 V at various current densities. The electrochemical impedance spectroscopy (EIS) measurement was analyzed within a frequency range of 100 kHz to 0.1 Hz at an open-circuit potential with an a.c. amplitude of 10 mV. Cyclic voltammetry (CV) and EIS were carried out with a Zahner electrochemical unit (1M6e, Zahner, Germany). Galvanostatic charge–discharge cycling of the cells for reliability study was performed with a battery tester (NAGANO, BTS-2004H, Japan).

**Results and Discussion**

**Characterization of synthesized graphene**

Figure 1 shows the SEM image of synthesized graphene. The surface morphology of the graphene appears to be highly porous, giving more access to the electrolyte. This might have high surface access even when they are stacked to electrode configuration. The nitrogen adsorption−desorption isotherm has been employed to study specific surface area (SSA) and porosity of the graphene. Figure 2 shows the BET isotherm of graphene at standard temperature and pressure (STP), which reveals a specific surface area of 437.77 m$^2$/g. The hysteresis between adsorption and desorption isotherm along with sharp fall in adsorbed amount, at higher relative pressures, can be assigned to the mesoporous nature of the graphene sheets. Figure 2b shows desorption pore volume as a function of pore radius dV/dlog(r), calculated using BJH method. It clearly shows the distribution of average pore size in diameter with maxima at 3.77 nm. Because the average pore size of the nanocomposite is higher than the dimension of the ionic liquid ions (~0.7 nm), it enables ions to accommodate inside the pores and thus results in better electrolyte accessibility and improves the charge storage. This mesoporous nature would allow electrolyte to access even the interior region of electrode when it is pressed as an electrode.

Figure 3 shows the X-ray diffraction pattern of the graphene. The broad peak at 24.8 ° corresponds to the (002) hexagonal plane of graphene. The d value 3.76 Å obtained from the fit into hexagonal is in well agreement with the graphitic nature of carbon. The XRD indicates the graphene synthesized is highly reduced from graphene oxide and no appearance of graphene oxide peak. The broad nature of the peak indicates disorder created during synthesis by modified Hummer's method. Figure 4 shows Raman spectra of graphene powder as prepared. Lorentzian fitting was done to obtain the positions and widths of the D and G bands in the Raman shift spectra. The Raman spectrum shows two broad peaks namely G-band and D-band which are characteristic for graphitic nature of carbon. The G-band at 1584 cm$^{-1}$ originates from ordered graphitized carbon while the D-band at 1350 cm$^{-1}$ is due to the disordered-activated band. The prominent peak at 1584 cm$^{-1}$ can be attributed to sp$^2$ bonded carbon atoms.[26] Moreover, the 2D peak at 2689 cm$^{-1}$ has appeared which confirms the crystalline nature in spite of defects introduced during the chemical modification.

Figure 5a shows the high-resolution TEM image of synthesized graphene. From the image, it is clearly seen that the prepared graphene is crippled and wrinkled in nature. Such structure helps in not restacking and prevents stacking of the graphene sheets together. The TEM image indicates the presence of single layer and highly porous nature of the synthesized graphene. The selected area electron diffraction (SAED) in Fig. 5b shows a ring like pattern consisting of many diffraction spots for each order of diffraction. This ring like spot is attributed to the hexagonal pattern with few graphene layers. From the TEM images it is clearly seen the sample prepared consists of graphene platelet from few layers to single layer. The wrinkle and cripple nature of the graphene helps for more access of the electrolyte during electrochemical activities.

The chemical composition of the as-prepared graphene was deduced from XPS measurement. As shown in Fig. 6, the full width at half maximum (FWHM) of the main sp$^2$ carbon peak of synthesized graphene is 1.18 eV, which is reasonably close to values in fits to highly oriented pyrolytic graphite.[7] Multiple states are also present on the high binding energy side of the main sp$^2$ peak. From the spectra, C1s peak is observed at 284.6 eV which indicates the sp$^2$ graphite in nature. This confirms the prepared graphene is highly reduced after the reduction process. But there are few shoulder peaks around 286 eV and 287 eV which are due to hydroxyl group C-OH group and carbonyl group C=O, respectively. Similarly there is a broad peak around 291 eV, which is due to carboxylic group presence. This indicates that the as-prepared graphene is not fully chemically reduced but the amount of functional group peaks is very low compared to C1s peak.

The vibrational characteristics of graphene were measured using Fourier transform infrared spectroscopy (FTIR). Figure 7 shows the FTIR spectra of graphene. The following bands were observed: O-H stretching (3200-3400 $cm^{-1}$), C=O and C-O stretching 1629 $cm^{-1}$ and aromatic C=C stretching (1400-1600 $cm^{-1}$). The peak at 3433 $cm^{-1}$ corresponds to stretching vibrations of hydroxyl group. The band corresponding to stretching antisymmetric and symmetric stretching vibrations of =$CH_2$ occur at 2925 and 2853 $cm^{-1}$, respectively, for graphene. This can be attributed to the presence of graphene and the residual oxygen containing functional groups in exfoliated graphene.

**Electrochemical characterization of the symmetric graphene supercapacitor**

To measure the electrochemical performance of the graphene, a two electrode cell geometry was prepared using EBIMF 1 M ionic liquid as electrolyte. The cyclic voltammetry measurements were carried out from 0 - 3.5 V at various scan rates from 5- 50 $mVs^{-1}$. Figure 8 shows the CV curve for graphene electrodes at different scan rates. From the graph, it is clearly seen that a rectangular shaped curve is observed for all scan rates, which indicates an electrical double layer capacitor of the electrode material with no pseudocapacitance effect due to functional groups. The specific capacitance measured from the rectangular CV curves was 204 $Fg^{-1}$ at a scanning rate of 5 mV $s^{-1}$ (2 A/g) between 0 - 3.5 V.

Figure 9 shows the charge and discharge curve of graphene electrode at a constant current of 2.5 $Ag^{-1}$. The slope variation of charge/discharge curves with respect to the time dependence of potential illustrates that the double layer capacitance behavior of the electrodes resulted from the electrochemical adsorption and desorption from the electrode-electrolyte interface. The shape of the charge/discharge curve is in typical triangular shape, which again indicates that there is no pseudo capacitance distortion behavior. It is also noted from the Fig. 9 that the charge and discharge time of cells were almost same suggesting 100% Columbic efficiency.

The galvanostatic discharge at a current density of 2.5 Ag$^{-1}$ resulted in a specific capacitance of 196 Fg$^{-1}$ based on the total weight of the electrode materials. This value agrees well with the specific capacitance measured from the cyclic voltammetry curves. The high specific capacitance might be due to more porosity and high surface area of the graphene synthesized. It is observed that the discharge capacitance was monotonically decreased with an increase in current density. This may be due to the low penetration of the ions into the inner region of pores due to fast potential changes. The supercapacitors exhibited appreciably-high specific capacitance even at a high current density. As the charging current rate increases from 2.5 A/g, 5 A/g, to 7.5 A/g, the specific capacitance value decreases from 196 Fg$^{-1}$, 150 Fg$^{-1}$ and 137 Fg$^{-1}$, respectively. The lower values of capacitance at high current rates might be due to less ionic penetration in the electrode surface compared to the case at a low current rate. As expected, the capacitance of the cell decreases linearly with increasing current densities, which is the typical behavior of electrochemical supercapacitors.[13,14] We have also characterized the cycle rate performance of the specific capacitance with respect to charge-discharge cycles. Figure 10 shows the specific capacitance for 10,000 cycles at a current density of 7.5 Ag$^{-1}$. The devices exhibited excellent stability and reliability at high current charge/discharge cycles. After 10000 cycles, as high as 98 % of its capacitance was still retained. The high stability of our supercapacitors at high current densities suggests that these energy storage devices are suitable for fast charging applications.

Electrochemical impedance spectroscopy study on the graphene electrodes is shown in Fig.11. The Nyquist plot of graphene based supercapacitor shows an inclined line in the low frequency region and a semicircle in the high frequency region. The straight line at low frequency indicates a nearly ideal capacitor response. If the inclined line is near to straight line it is related to ideal capacitor behavior.[27,28] From Fig. 11, the equivalent series resistance (ESR) value obtained from the x-intercept of the Nyquist plot is 20 Ω. The Nyquist plot of graphene supercapacitor after cycling is also shown in Fig. 11. The ESR value is increased to 25 Ω after 30,000 cycles under various current rates namely, 2.5 Ag$^{-1}$, 5 Ag$^{-1}$ and 7.5 Ag$^{-1}$ with 10,000 cycles at each rates. The increase in ESR value may be due to the slight electrode expansion after multiple cycling.

The Ragone plot of the symmetric graphene supercapacitor is shown in Fig. 12. The energy density and power density are calculated using the method reported.[6] The energy density values were 83.36, 64.18, 58.25 Whkg$^{-1}$ at current density values of 2.5, 5, 7.5 Ag$^{-1}$, corresponding to discharge time 69 sec, 25 sec, 16 sec, respectively. These values are quite low compared to the previously reported values. Especially, the charge/discharge time is very less compared to previously reported values so far.[4] The time required to charge 64.18 Whkg$^{-1}$ is around 25 seconds whereas in the previous reports[4] the similar value around 67.7 Whkg$^{-1}$ was achieved in 50 seconds for graphene based supercapacitors. Here the discharge time has reduced to almost half of the time for a similar value of energy density reported earlier. Apart from the superior properties of synthesized graphene, the reduced charge/discharge time could be attributed to the more efficient current collecting method we have employed in this work such as using SS mesh and Ketchen black conductive additive in cell assembly. The power density obtained at current density 7.5 Ag$^{-1}$ is 13.12 kW kg$^{-1}$ with an energy density of 58.25 Whkg$^{-1}$, thus proving a possible adoption for electric vehicle applications.[29, 30] According to our knowledge, these values are the highest so far reported in the literature. Further, this method could be scaled to large scale manufacturing as the method developed here could be realized by roll to roll technology where the active materials slurry can be applied uniformly to SS mesh and can be rolled in cylindrical shape.

**Conclusions**

In summary, a high energy density, power density and specific capacitance of 64.18 Whkg$^{-1}$, 8.75 kWkg$^{-1}$ and 150.9 Fg$^{-1}$ at current density 5 Ag$^{-1}$ were realized by highly porous graphene based supercapacitors. The retentivity of the capacitance after several tens of thousands of cycling is stable. The porous nature of the graphene and highly reduced graphene enhance the accessibility for ion diffusion and high conduction. The supercapacitor energy storage devices with high specific capacitance and short charging time demonstrated here can be scaled up for manufacturing in the near future for electric vehicle applications.

**Acknowledgement:** This work was supported by the World Class University (WCU) program at GIST through a grant provided by the Ministry of Education, Science and Technology (MEST) of Korea.

**Broader context**

The demand for supercapacitors is growing for most of consumer electronics as well as industrial applications. Their rapid charging and power delivery capabilities find vast applications with dynamic load variations. In many real time applications, there is a critical need that supercapacitors with high energy storage density can be quickly charged/discharged at high current rates. Graphene or other active carbon based supercapacitors have demonstrated great potential for energy storage applications but so far there has bee little work reported on supercapacitors that can operate at high current ratings with cycle life stability. In this work we investigate graphene-based supercapacitors at high current densities for long durability. We show supercapacitors stable operation after several tens of thousands of cycles at high current. The devices exhibit a specific capacitance at high current that is comparable to the value reported so far at a very low current level.

**Notes and References**

**Figure Captions**

Figure 1. Typical SEM image of as-prepared graphene.

Figure 2. (a) Nitrogen adsorption isotherm and (b) pore size distribution of the graphene.

Figure 3. X-ray diffraction pattern of the graphene powder.

Figure 4. (a) HR TEM image of the graphene few layers. Inset shows the corresponding selective area electron diffraction pattern. (b) Wrinkles and crumpling of graphene sheet.

Figure 5. shows the XPS C1s spectra of graphene.

Figure 6. Raman spectra of graphene.

Figure 7. FTIR spectra of as prepared graphene.

Figure 8. C-V curve of graphene supercapacitor.

Figure 9. Charge/discharge curve for graphene supercapacitor for various current densities, namely, 2.5 $Ag^{-1}$, 5 $Ag^{-1}$, 7.5 $Ag^{-1}$.

Figure 10. Cycle rate performance of graphene supercapacitor at a current density of 7.5 $Ag^{-1}$.

Figure 11. Electrochemical impedance spectra of graphene supercapacitor.

Figure 12. Ragone plot of graphene supercapacitor at various charge-discharge rates.

Figure 1   Typical  SEM image of as-prepared graphene.

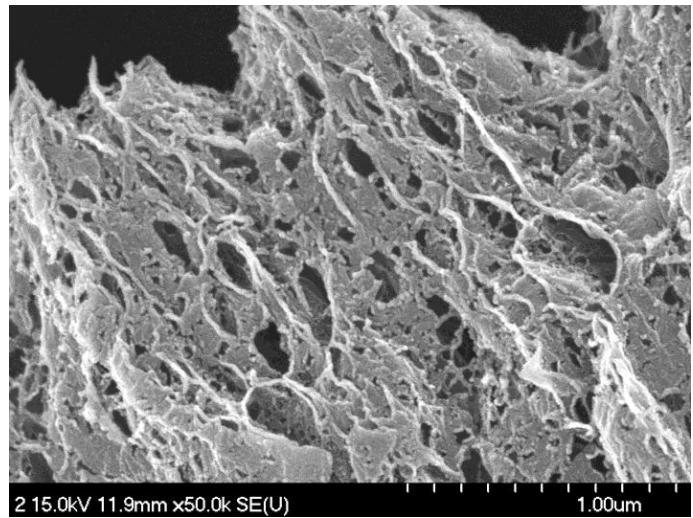

Figure 2 (a) Nitrogen adsorption isotherm and (b) pore size distribution of the graphene.

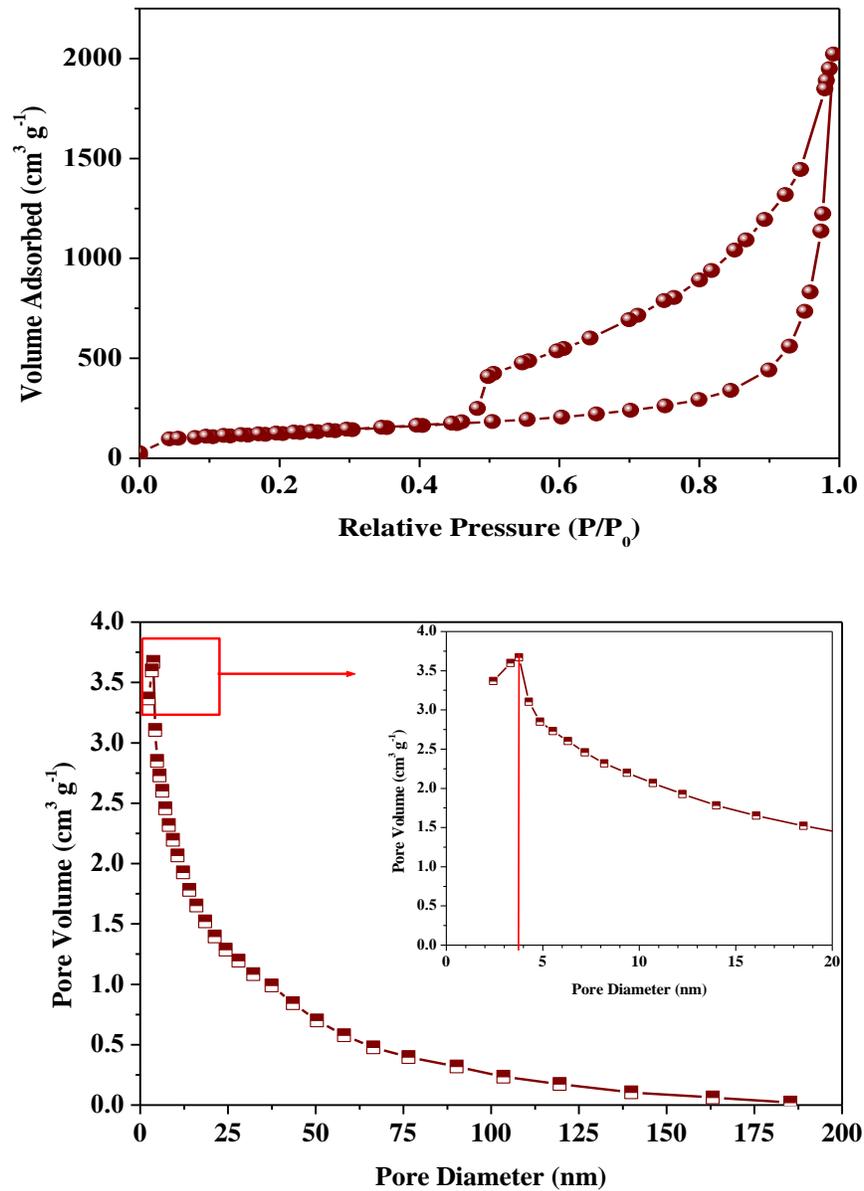

Figure 3  X-ray diffraction pattern of the graphene powder.

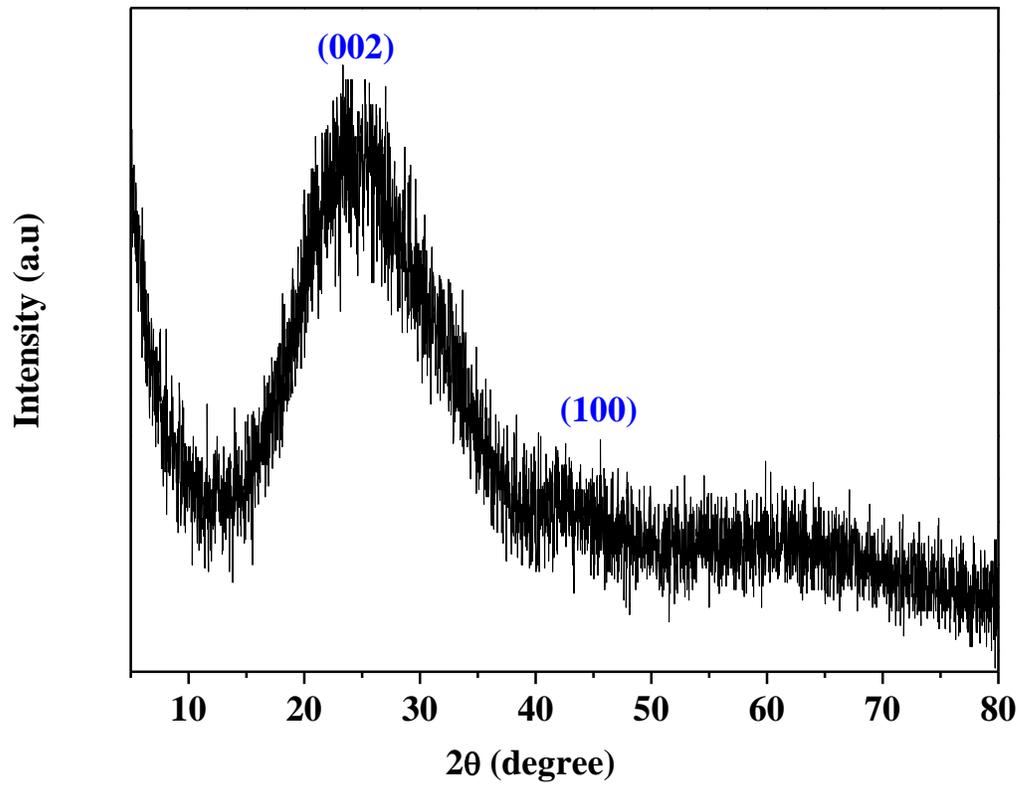

Figure 4  Raman spectra of graphene.

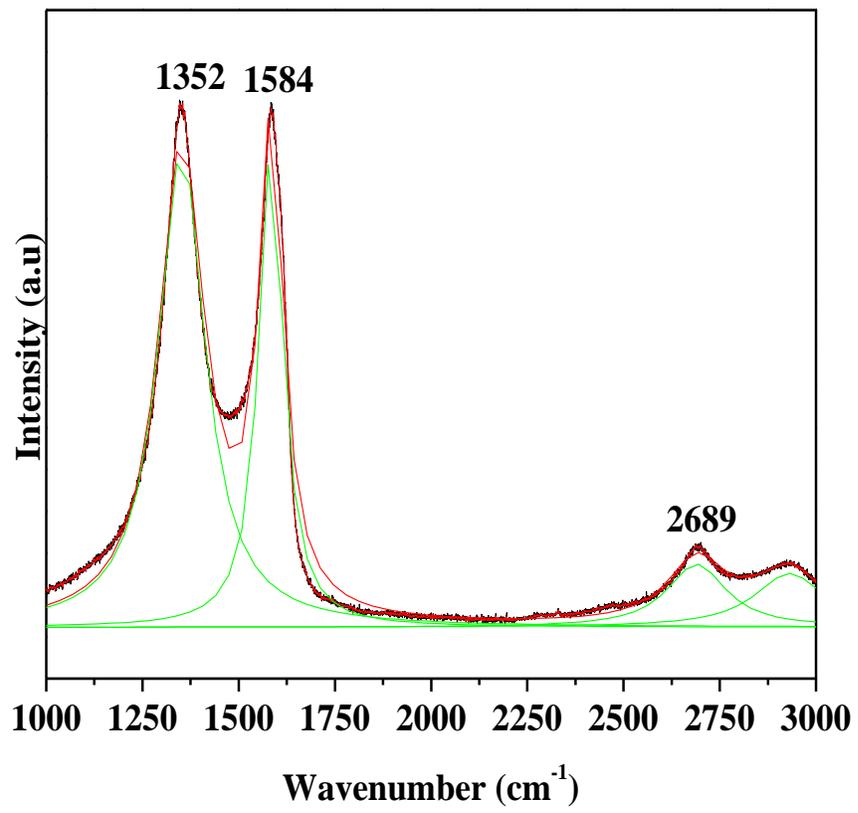

Figure 5 (a) Wrinkles and crumpling of graphene sheet. (b) HR TEM image of the graphene few layers. Inset shows the corresponding selective area electron diffraction pattern.

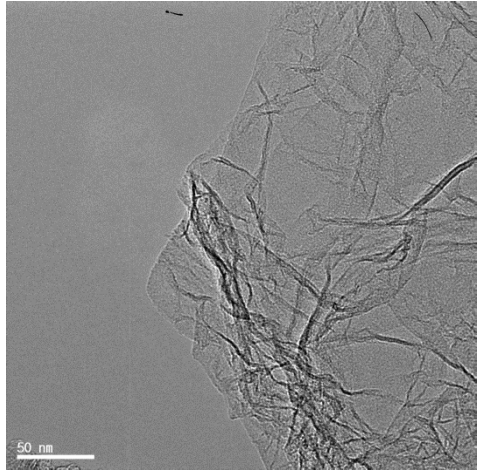

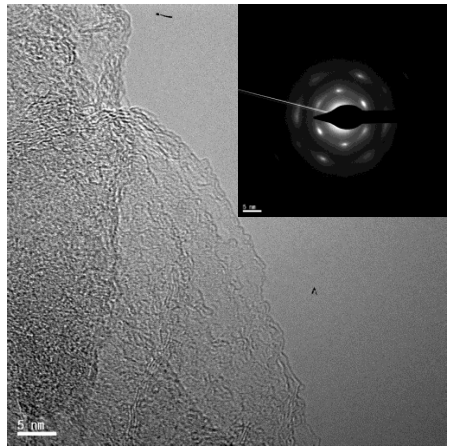

Figure 6 shows the XPS C1s spectra of graphene.

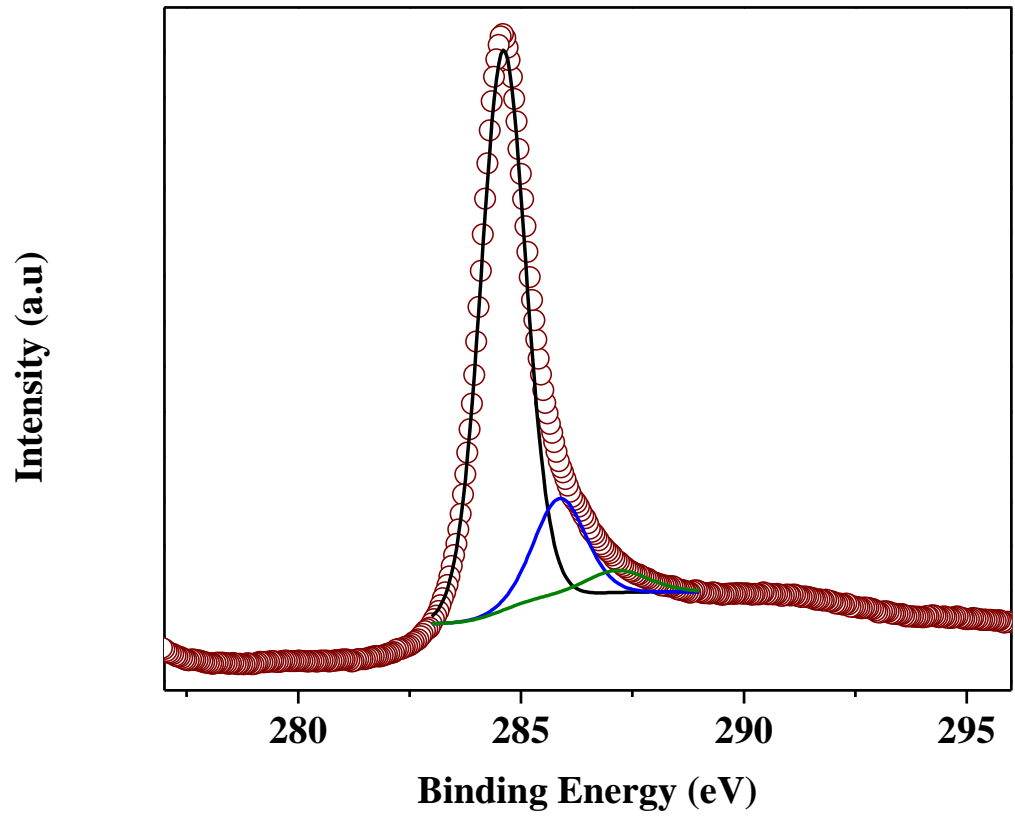

Figure 7 FTIR spectra of as prepared graphene.

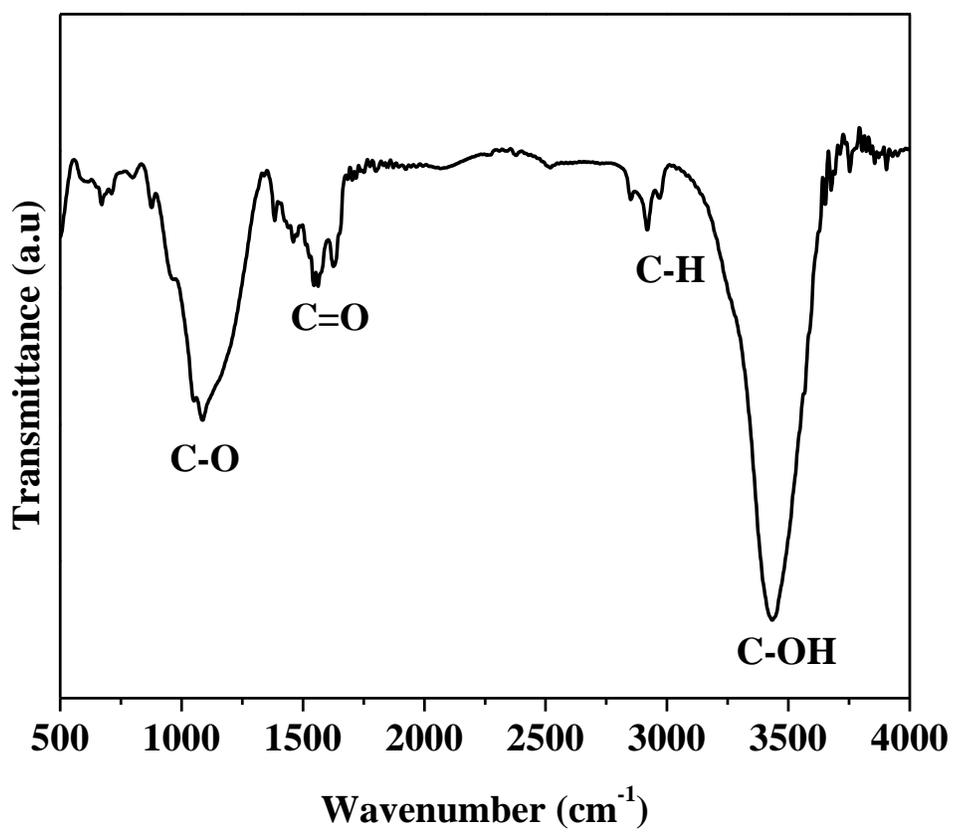

Figure 8 C-V curves of graphene supercapacitor.

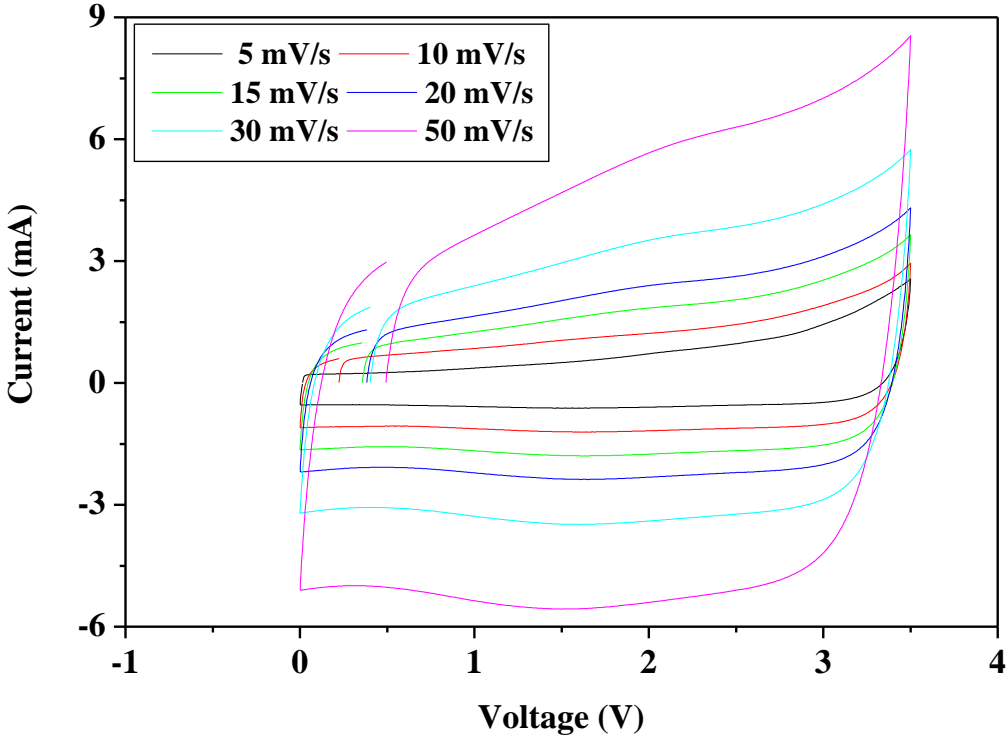

Figure 9 Charge/discharge curve for graphene supercapacitor at various current density 2.5 Ag$^{-1}$, 5 Ag$^{-1}$, 7.5 Ag$^{-1}$.

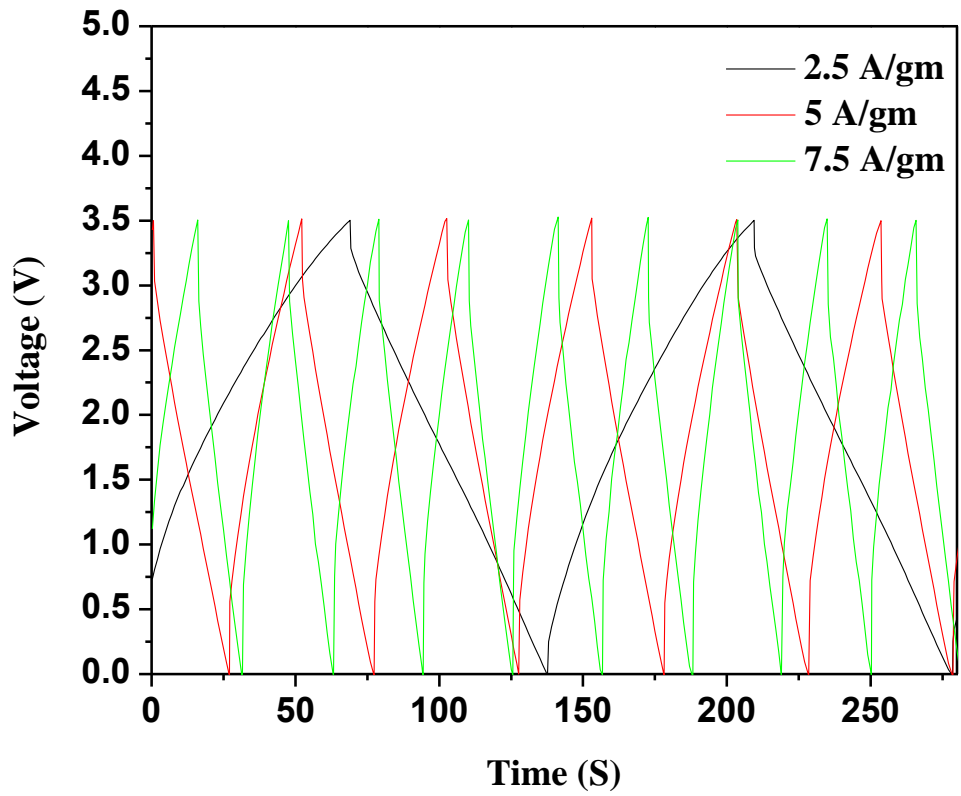

Figure 10 Cycle performance of graphene supercapacitor at a current density of 7.5 Ag⁻¹.

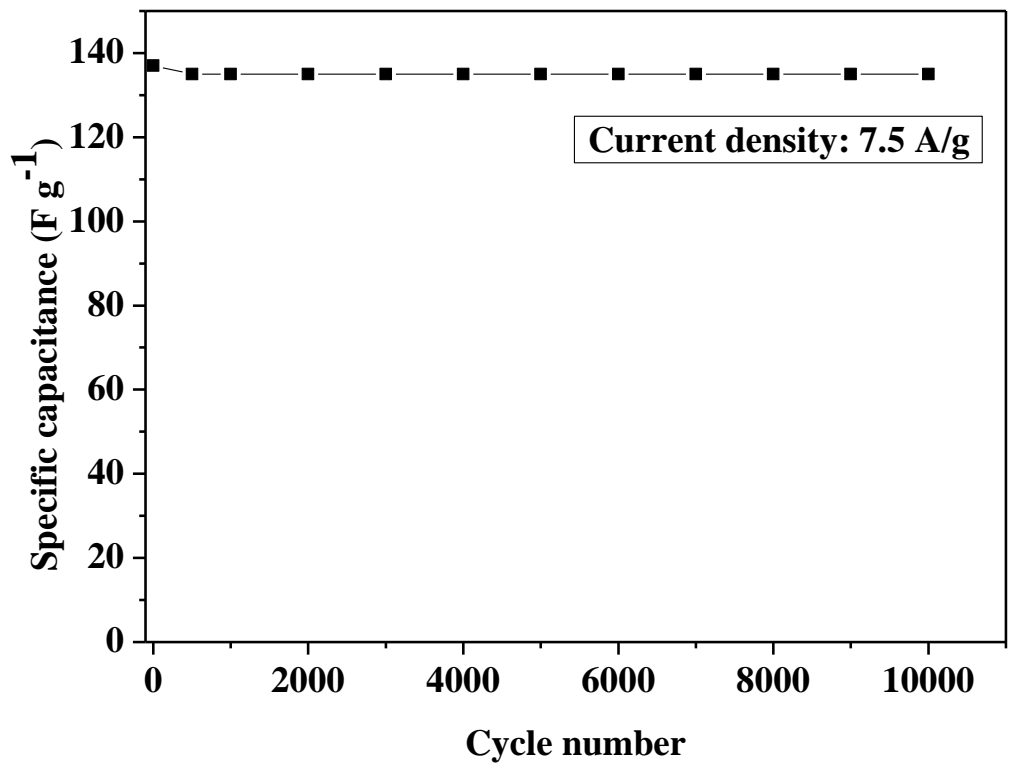

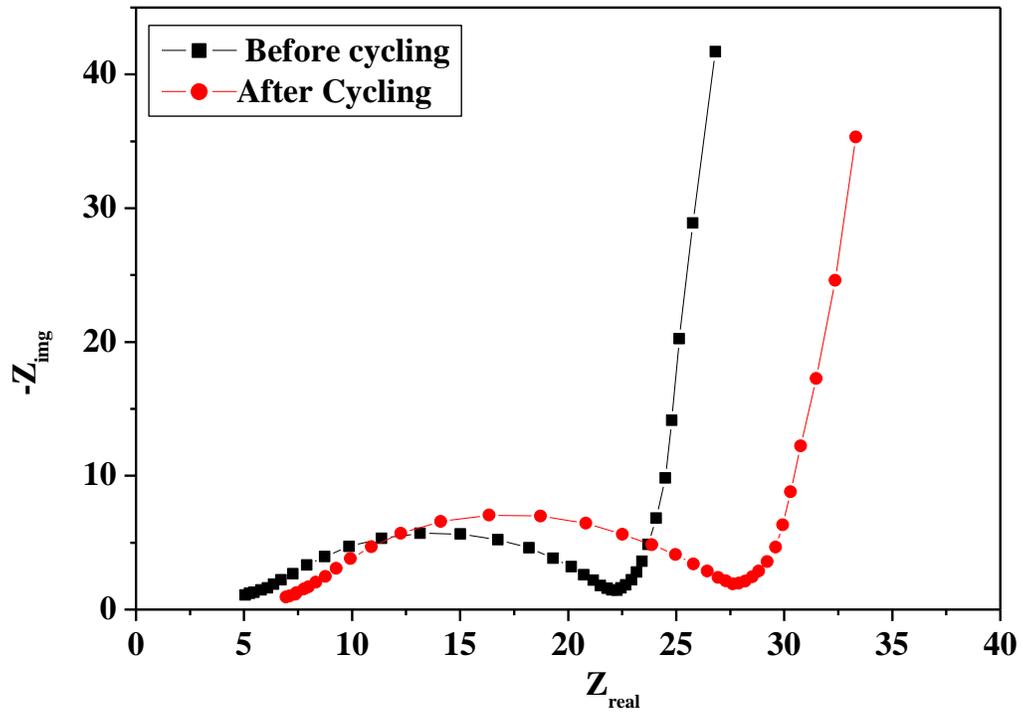

Figure 11 Electrochemical impedance spectra of graphene supercapacitor.

Fig.12 Ragone plot of graphene supercapacitor at various charge-discharge rates.

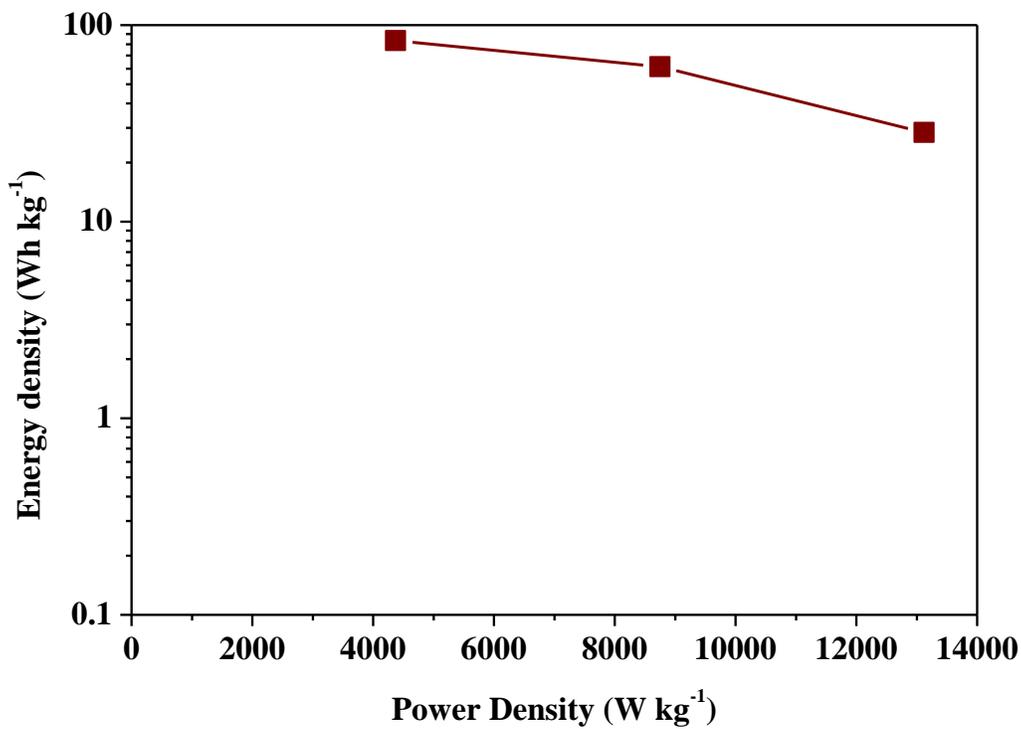